\documentclass[prl,twocolumn,floatfix,groupedaddress]{revtex4}
\usepackage{graphicx}
\usepackage{sansmath}
\usepackage[caption=false]{subfig}
\usepackage{bm}
\usepackage{mathtools}
\usepackage{tabularx}
\newcolumntype{Y}{>{\centering\arraybackslash}X}

\newlength{\onecolfig}
\newlength{\twocolfig}
\newcommand{\ion}[2]{\mbox{$^{#2}$#1$^+$}}
\newcommand{\Ca}[1]{\ion{Ca}{#1}}

\newcommand{\Mg}[1]{\ion{Mg}{#1}}

\newcommand{\unit}[1]{\,\mbox{#1}}

\newcommand{\Hz}{\unit{Hz}}
\newcommand{\kHz}{\unit{kHz}}
\newcommand{\MHz}{\unit{MHz}}
\newcommand{\GHz}{\unit{GHz}}

\newcommand{\uW}{\unit{$\mu$W}}

\newcommand{\um}{\unit{$\mu$m}}

\newcommand{\s}{\unit{s}}

\newcommand{\ms}{\unit{ms}}
\newcommand{\us}{\unit{$\mu$s}}

\newcommand{\mT}{\unit{mT}}

\newcommand{\etal}{{\em et al.}}

\newcommand{\ish}{\mbox{$\sim$}\,}
\newcommand{\ltish}{\protect\raisebox{-0.4ex}{$\,\stackrel{<}{\scriptstyle\sim}\,$}}

\newcommand{\up}{\mbox{$\uparrow$}}
\newcommand{\down}{\mbox{$\downarrow$}}

\newcommand{\ket}[1]{\mbox{$\left| #1 \right>$}}

\newcommand{\wee}[2]{\mbox{$\frac{#1}{#2}$}}

\newcommand{\MS}{M{\o}lmer-S{\o}rensen}

\begin{document}
\bibliographystyle{apsrev}

\title{High-fidelity trapped-ion quantum logic using near-field microwaves} 

\author{T. P. Harty, M. A. Sepiol, D. T. C. Allcock\footnote{Present address: National Institute of Standards and Technology, 325 Broadway, Boulder, CO 80305, U.S.A.}, C. J. Ballance, J. E. Tarlton and D. M. Lucas}
\affiliation{Department of Physics, University of Oxford, Clarendon Laboratory, Parks Road, Oxford OX1 3PU, U.K.}
\date{27 June 2016}

\begin{abstract}
We demonstrate a two-qubit logic gate driven by near-field microwaves in a room-temperature microfabricated ion trap. We measure a gate fidelity of 99.7(1)\%, which is above the minimum threshold required for fault-tolerant quantum computing. The gate is applied directly to \Ca{43} ``atomic clock'' qubits (coherence time $T_2^*\approx 50\,\mathrm{s}$) using the microwave magnetic field gradient produced by a trap electrode. We introduce a dynamically-decoupled gate method, which stabilizes the qubits against fluctuating a.c.\ Zeeman shifts and avoids the need to null the microwave field.
\end{abstract}

\maketitle

Laser-cooled trapped atomic ions are a promising platform for the development of a general-purpose quantum computer~\cite{Monroe2013}. In common with other technologies, the present challenge is performing all elementary logic operations with the fidelity necessary for quantum error correction, whilst using techniques which can be scaled to the number of qubits required to perform a useful computation. Trapped-ion qubits are based on either optical~\cite{Nagerl2000} or hyperfine~\cite{Monroe1995} atomic transitions. Hyperfine qubits lie in the convenient microwave domain, and have exhibited minute-long memory coherence times~\cite{Bollinger1991,Harty2014}. Nevertheless, they are usually manipulated via optical (Raman) transitions, firstly because of the convenience of addressing individual ions with tightly-focussed laser beams~\cite{Nagerl1999}, and secondly because the short optical wavelength allows efficient multi-qubit logic gates  based on coupling the ions' spin and motional degrees of freedom~\cite{Wineland1998}.

Microwave methods have been proposed~\cite{Mintert2001,Ospelkaus2008,Timoney2011}, and recently demonstrated, both for individual qubit addressing~\cite{Warring2013,Piltz2015,AudeCraik2016}, and for multi-qubit logic gates~\cite{Ospelkaus2011,Khromova2012,Weidt2016}. This offers the prospect of performing all coherent operations using purely electronic methods, making phase control significantly easier, and replacing lasers with cheaper, smaller, more stable microwave devices. Microwave elements can also be integrated into trapping structures more easily than their optical counterparts for improved scalability~\cite{Allcock2013}. Furthermore, microwave gates can theoretically attain higher fidelities as they are not fundamentally limited by photon scattering~\cite{Ozeri2007}. Two distinct microwave methods are being pursued: using far-field microwaves in combination with a local static magnetic field gradient; and, using a local near-field microwave magnetic field gradient. Microwave-driven two-qubit gates have previously been reported in a single experiment using the near-field method (with 76\% fidelity~\cite{Ospelkaus2011}), and in two experiments using the far-field method (with 70\% fidelity in a 3-ion chain and, very recently, 98.5\% for a pair of ions~\cite{Khromova2012,Weidt2016}). Beyond quantum information processing, microwave quantum logic techniques are also applicable to metrology and high-resolution spectroscopy, for example for the study of systems without accessible optical transitions~\cite{Wineland1998,Niemann2014}.

In this Letter, we report a near-field microwave two-qubit gate with fidelity exceeding the $\approx 99\%$ minimum threshold required for fault-tolerant quantum error correction~\cite{Fowler2012}, and comparable to that of the best reported fidelities achieved with lasers or other qubit technologies~\cite{Benhelm2008,Ryan2009,Barends2014,Ballance2016,Tan2016}. We estimate the main sources of error in the gate. We also set a limit on the errors induced by the gate fields on an ``idle'' memory qubit. The two-qubit gate operation is implemented with the same qubit states, and in the same device, which we have previously used in demonstrating high-fidelity ($>99.9\%$) single-qubit state preparation, gates, memory and readout~\cite{Harty2014}. The trap is a lithographically-defined two-dimensional surface-electrode design, incorporating integrated microwave waveguides and resonators, and is operated at room temperature~\cite{Allcock2013}. Surface traps are especially promising for scaling up to large numbers of trap zones, as proposed for a ``quantum CCD'' architecture~\cite{Wineland1998}.

\begin{figure}[t]
\includegraphics[width=\columnwidth]{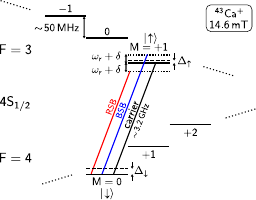}
\caption{Part of the ground-level hyperfine structure of \Ca{43} at 14.6\mT\ (not to scale), showing the clock-qubit states (\ket{\downarrow} and \ket{\uparrow}), other states which are connected to the qubit by spectator microwave transitions, and the microwave fields used for the gate. The blue and red sideband fields (BSB and RSB) have frequencies \mbox{$(\omega_0+\Delta)\pm(\omega_{\mathrm{r}}+\delta)$}, where $\omega_0$ is the unperturbed qubit frequency, $\omega_{\mathrm{r}}$ is the motional mode frequency, $\delta$ is the gate detuning, and $\Delta=\Delta_{\uparrow}-\Delta_{\downarrow}$ is the differential a.c.\ Zeeman shift produced by the (strong) sideband fields. The (weak) carrier field used for dynamical decoupling is tuned to resonance with the shifted qubit transition at $(\omega_0+\Delta)$.}
\label{fig:MS_Levels}
\end{figure}

This work was performed using the \Ca{43} intermediate-field atomic clock qubit described in \cite{Harty2014}. The qubit is formed from a pair of hyperfine states in the ground-level, separated by a 3.200\GHz\ transition (figure~\ref{fig:MS_Levels}), whose frequency is first-order independent of magnetic field at a static field of 14.6\mT. Details of the laser cooling, initialisation and measurement of this qubit may be found in \cite{Harty2014,Allcock2016}.

The two-qubit gate implemented in this work is an extension of the ideas of M{\o}lmer and S{\o}rensen (MS), Ospelkaus \etal, and Bermudez \etal~\cite{Sorensen2000,Ospelkaus2008,Bermudez2012}; it is a gate driven by a microwave near-field gradient, which is robust to what would otherwise be the largest source of experimental error in our system, {\em viz.}\ fluctuating a.c.\ Zeeman shifts arising from the microwave fields. A standard MS gate is implemented with a bichromatic field with frequencies near the first red and blue sideband transitions for one of the ions' normal modes of motion, resulting in dynamics described by the Hamiltonian
\begin{equation}
H_{\mathrm{MS}} = \wee{1}{2}\hbar\Omega S \left(a e^{\mathrm{i}\delta t} + a^{\dagger}e^{-\mathrm{i}\delta t}\right)
\label{eq:H_MS}
\end{equation} 
Here, $\Omega$ and $\delta$ are the gate Rabi frequency and detuning respectively (see figure \ref{fig:MS_Levels}), and \mbox{$S=\sigma_{x,1} \pm \sigma_{x,2}$} is the collective spin operator, where $\sigma_{x,i}$ is the Pauli operator acting on ion $i$ and the sign is positive (negative) if the ions' normal motions are in phase (anti-phase).

\begin{figure*}[t]
\captionsetup[subfigure]{labelformat=empty}
\subfloat[]{\includegraphics[width=\columnwidth]{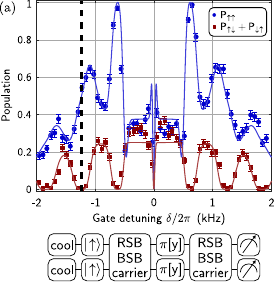}} \hfill
\subfloat[]{\includegraphics[width=\columnwidth]{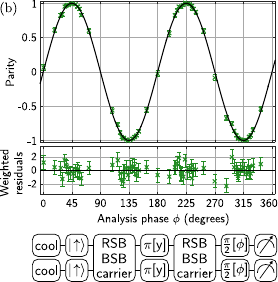}}
\caption{Experimental demonstration of a four-loop DDMS gate incorporating continuous dynamical decoupling and refocussing $\pi[y]$-pulse. 
(a) State populations as a function of sideband detuning $\delta$. Solid lines show a numerical simulation for the ideal dynamics, starting from a ground-state cooled motional mode. The dashed line indicates the detuning used for the entangling gate, whose total duration is $t_g=8\pi/\left|\delta\right|=3.25\ms$. 
(b) Measurement of the parity, defined as $\mathrm{P}_{\uparrow\uparrow}+\mathrm{P}_{\downarrow\downarrow}-\mathrm{P}_{\uparrow\downarrow}-\mathrm{P}_{\downarrow\uparrow}$, used to determine the fidelity of the Bell state $\ket{\Psi}=\ket{\uparrow\uparrow}+i\ket{\downarrow\downarrow}$ produced by the gate. The data consist of five separate experimental runs, which were interleaved with measurements of the SPAM error and Bell state populations. A maximum-likelihood fit (solid line), assuming binomial statistics~\cite{Ballance2016}, gives a parity amplitude of $0.9953(23)$. The phase offset is determined from an independent calibration and is not floated in the fit. Error bars represent $68\%$ confidence intervals; all data have been corrected for SPAM errors (see text).}
\label{fig:results}
\end{figure*}

The ions' motion is driven by the spatial gradient of the microwave magnetic field. In general, this gradient will be accompanied by a non-zero field amplitude at the ions' equilibrium positions. (The field can be made small by nulling with additional microwave electrodes, or with specific trap geometries~\cite{Carsjens2014}, but in practice it will always be present at some level.) This unwanted field will drive off-resonant Rabi flopping (spin flips) on any hyperfine transition connecting to the qubit states, but the effect of this on the operation of the gate can be highly suppressed using pulse-shaping techniques~\cite{Roos2008}. The field also induces a differential a.c.\ Zeeman shift $\Delta$ on the qubit transition (figure~\ref{fig:MS_Levels}), described by the Hamiltonian
\begin{equation}
H_{\mathrm{Z}} = \wee{1}{2}\hbar\Delta\left(\sigma_{z,1}+\sigma_{z,2}\right)
\end{equation}
A constant a.c.\ Zeeman shift may be treated as an effective change in the qubit frequency and compensated for by adjusting the microwave frequencies appropriately. However, any fluctuations in $\Delta$ will lead to qubit dephasing, which can be a significant source of error. Henceforth, we assume that the bulk of the a.c.\ Zeeman shift has been compensated, and use $\Delta'$ to represent the residual fluctuations (which we take to be slowly varying compared with the gate's duration).

If $H_{\mathrm{MS}}$ acted in the $\sigma_z$ basis, it would commute with $H_\mathrm{Z}$ and this error could be suppressed by performing the gate inside a spin-echo sequence~\cite{Leibfried2003}. However, $\sigma_{z}$ gates are not straightforward to implement with microwaves~\cite{Ospelkaus2008}. Instead, we take advantage of the fact that $H_{\mathrm{MS}}$ does commute with a carrier drive of the qubit transition, provided that the carrier phase is chosen to produce rotations about the same axis of the Bloch sphere as $H_{\mathrm{MS}}$. The corresponding Hamiltonian is
\begin{equation}
H_{\mathrm{c}}= \wee{1}{2}\hbar\Omega_{\mathrm{c}} \left( \sigma_{x,1} + \sigma_{x,2}  \right)
\end{equation}
This carrier drive dynamically decouples the qubit from $H_{\mathrm{Z}}$, as can be seen by considering the total system Hamiltonian, \mbox{$H_{\mathrm{T}} = H_{\mathrm{MS}} + H_{\mathrm{c}} + H_{\mathrm{Z}}$}, in the interaction picture with respect to $H_{\mathrm{c}}$:
\begin{equation}
H_{\mathrm{I}} = H_{\mathrm{MS}} + \wee{1}{2}\hbar\Delta'\sum\limits_{i=1,2}\sigma_{z,i} \cos{\Omega_{\mathrm{c}}t} + \sigma_{y,i} \sin{\Omega_{\mathrm{c}}t}
\label{eq:MS_rot_frame}
\end{equation}
If $\Omega_{\mathrm{c}} \gg \Omega, \Delta'$ the summed terms in \eqref{eq:MS_rot_frame} oscillate rapidly and may be disregarded. The rotating-frame Hamiltonian then reduces to $H_{\mathrm{MS}}$. Furthermore, setting $\Omega_{\mathrm{c}}t_{\mathrm{g}} = 2m\pi$ for gate time $t_{\mathrm{g}}$ and integer $m$ ensures that the rotating frame coincides with the lab frame at $t_{\mathrm{g}}$, so that an error-free MS gate is achieved in both frames.

The requirement that \mbox{$\Omega_{\mathrm{c}} t_{\mathrm{g}} = 2m\pi$} may be avoided by using a composite gate sequence, with an additional $\pi$-pulse on each ion mid-way through the gate to refocus any partially complete carrier Rabi oscillations~\cite{Tan2013,Lemmer2013} (figure \ref{fig:results}). For this to work, the gate must be composed of an even number of phase-space loops, so that the $\pi$-pulse is applied while the ions' spins are disentangled from their motion. In this case, the gate is not sensitive to the exact value of $\Omega_{\mathrm{c}}$, provided that the applied pulse area is the same for each half. This sequence has the added benefit of being insensitive to transient a.c.\ Zeeman shifts at the beginning and end of each each half. Moreover, if the $\pi$-pulse phase is chosen to give a rotation about the $y$-axis, errors due to drifts in the motional mode frequency are also suppressed~\cite{Hayes2012}.

The dynamically-decoupled MS (DDMS) gate described above is closely related to the ``single-sideband'' (SSB) gate proposed and demonstrated in~\cite{Bermudez2012,Lemmer2013,Tan2013}, which uses only one of the red or blue sideband fields in combination with a carrier drive. The SSB gate was originally introduced for use with lasers, where it has the advantage that, unlike the MS gate, it does not require interferometric stability between optical fields. This advantage is inconsequential for microwave gates due to the relative ease of accurately controlling microwave phases.

In their original proposal for the SSB gate, the authors noted that their carrier drive technique could be extended to the standard MS gate~\cite{Bermudez2012}. Our work develops this idea, identifying the importance of the relationship between the carrier and sideband phases (which is not significant for the SSB gate), and providing numerical modelling (Supplementary Material), as well as an experimental demonstration. The principle advantage of the DDMS gate is that, unlike the SSB gate and other ``dressed-state'' schemes~\cite{Bermudez2012,Zheng2014,Weidt2016}, the carrier drive is merely used to suppress noise, rather than forming a fundamental part of the gate mechanism. As a result, when $\Delta'=0$, the DDMS gate exactly reproduces the MS dynamics at all times and for all values of $\Omega_{\mathrm{c}}$. This is not true for the SSB gate, which is consequently very susceptible to noise in $\Omega_\mathrm{c}$ (see Supplementary Material). This is a significant limitation of the SSB gate, potentially requiring the use of second-order driving fields to achieve high fidelities~\cite{Lemmer2013}. Additionally, the DDMS gate requires half the total microwave power to achieve a given gate speed, reducing the power dissipated in the ion trap chip.


\begin{figure}[t]
\includegraphics[width=\columnwidth]{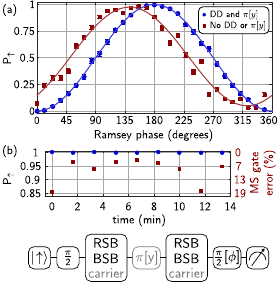}
\caption{%
Single-ion Ramsey experiment used to measure a.c.\ Zeeman shift fluctuations. For the blue data set (circles), we enclose the gate sequence in a pair of $\pi/2$-pulses, leaving all microwave frequencies and pulse durations set as for the two-qubit gate. For the red dataset (squares), we repeat the measurement without the dynamical decoupling or the refocussing $\pi$-pulse. (a) We scan the phase $\phi$ of the second $\pi/2$-pulse to obtain a fringe. Lines show maximum-likelihood fits to the data, giving fringe contrasts of $0.998(5)$ and $0.924(6)$. 
(b) We set $\phi$ to the fringe's peak and monitor fluctuations in $P_{\uparrow}$ over time. The average of the blue data points gives $P_{\uparrow}=0.9994(4)$, indicating the gate fields induce $\ltish 0.1\%$ loss of qubit coherence. The right-hand ordinate gives, for the red points, the simulated MS gate error corresponding to a given fringe contrast, assuming normally-distributed  shot-to-shot a.c.\ Zeeman shift fluctuations. All data have been corrected for SPAM errors.
}
\label{fig:AC_Zeeman_noise}
\end{figure}

For the experimental implementation, we confine a pair of \Ca{43} ions 75\um\ above the surface-electrode ion trap described in \cite{Allcock2013}. We perform the gate on one of the ions' radial rocking (out-of-phase) modes, whose secular frequency is 3.255\MHz~\footnote{The other radial rocking mode is at 3.588\MHz.}. This mode was chosen over the 3.286\MHz\ centre of mass (COM) mode because of its lower heating rate, which was measured to be $\lesssim5\,\mathrm{quanta/s}$ (compared with $60(15)\,\mathrm{quanta/s}$ for the COM mode). We generate the Paul trap radiofrequency drive (38.3\MHz, $\approx60\unit{V}$ amplitude) using a home-built high-stability source, which reduces fluctuations in the motional mode frequency to $\sim 30\Hz$ r.m.s. We further suppress errors due to the residual mode frequency fluctuations by cooling the rocking mode close to its ground state with Raman sideband cooling~\cite{Allcock2016}. We also ground-state cool the spectator rocking mode to minimise dephasing due to cross-phase modulation~\cite{Roos2008b}\footnote{Note that, neglecting this effect, gate errors due to finite temperature are negligible in the absence of detuning errors due to the extremely low microwave effective Lamb-Dicke parameter~\cite{Ospelkaus2008}.}. In future experiments, ground-state cooling could be achieved using microwave sideband cooling~\cite{Ospelkaus2011}.

We generate microwaves by upconverting r.f.\ at $\sim$\,300\MHz\ from a commercial direct digital synthesis (DDS) source~\footnote{The RF source is a Milldown board, manufactured by Enterpoint Ltd, featuring 4 channels of Analog Devices AD9910 DDS; upconversion is done with Eclipse Microwave Inc.\ IQ2040MP4 IQ mixers, fed with a $3.5\GHz$ LO.}. The RSB and BSB are generated and amplified separately, before being combined on a quadrature hybrid. After the hybrid, a custom cavity filter~\footnote{API Technologies Corp.\ C3200-30-3SS.} removes noise (which had been observed to excite microwave spectator transitions during the gate) from the signal before it is fed to one of the trap's microwave electrodes. To minimise the effect of off-resonant spin-flips, we turn the RSB and BSB on/off adiabatically with a rise/fall time of 3\us. Additionally, we pre-distort the sideband pulse envelope to compensate for slow ($\sim\ms$) power transients during the gate. We use a slow digital feedback loop based on an IC power detector to reduce long-term drifts in the sideband power. Finally, we apply a 200\Hz\ zero-peak linear ramp to the RSB and BSB DDS frequencies during the gate to compensate for an observed ``chirp'' in the radial mode frequency (which may originate from thermal transients in the trap caused by the microwaves). Using 2\unit{W} in each sideband, we achieve a gate Rabi frequency of $\Omega/2\pi \simeq 308\Hz$. The resulting differential a.c.\ Zeeman shift is $\Delta/2\pi=20.78\kHz$. For the carrier drive, we apply 3\uW\ to a second trap electrode, giving $\Omega_{\mathrm{c}}/2\pi=3.7\kHz$. Further details about the microwave fields are given in the Supplement. 


The gate sequence is shown in figure~\ref{fig:results}(a). The gate consists of 4 loops in motional phase-space, with a $y$-axis $\pi$-pulse ($3.2\us$ duration) mid-way through. The total gate time is $3.25\ms$. We measure the fidelity of the Bell state produced by the gate using the standard tomographic method described in~\cite{Leibfried2003}. The population after the gate is $P_{\downarrow\downarrow}+P_{\uparrow\uparrow}=0.9980(8)$. Combining this with the parity measurement shown in figure~\ref{fig:results}(b), we calculate a fidelity of 99.7(1)\%. Here, we have corrected for the independently-measured state-preparation and measurement (SPAM) error of 0.34(3)\% per qubit~\cite{Ballance2016}. A significant source of SPAM error, not present in~\cite{Harty2014}, is two-ion fluorescence detection using a photomultiplier, which could be significantly reduced using a camera~\cite{Burrell2010}.

To estimate the benefit of the DDMS gate over the basic MS scheme, we perform the single-ion Ramsey experiment shown in figure~\ref{fig:AC_Zeeman_noise}. Here, the RSB and BSB are set up as for a gate on the two-ion rocking mode, leaving them $\approx30\kHz$ detuned from the nearest single-ion motional mode. As a result, they create an a.c.\ Zeeman shift without coupling to the ion's motion. The fluctuations in this a.c.\ Zeeman shift are determined from the resulting loss of fringe contrast. Without the dynamical decoupling and refocussing pulse, we measure a fringe contrast of 0.924(6). Assuming normally-distributed shot-to-shot fluctuations in the a.c.\ Zeeman shift, this corresponds to $\Delta'=19.7(8)\Hz$ r.m.s., which would give a MS gate error of 5.6(5)\% (figure 3(b)). With the dynamical decoupling and refocussing pulse, we find no loss of fringe contrast at the level of the measurement's sensitivity. This experiment also implies that the DDMS gate fields would introduce $\ltish 0.1\%$ error on ``idle'' memory qubits stored in neighbouring trap zones in a multi-zone architecture.

The measured two-qubit gate error is consistent with the $\lesssim0.2\%$ error expected from the rocking mode heating rate and the $\sim0.2\%$ error expected from the independently-measured $\sim30\Hz$\ r.m.s.\ fluctuations in the rocking-mode frequency. We infer from the data in figure~\ref{fig:AC_Zeeman_noise} that errors due to off-resonant excitation and a.c.\ Zeeman shift contribute $\lesssim 0.1\%$ error. Similarly, from the agreement between theory and data in figure~\ref{fig:results}(a), we estimate the error due to systematic mis-calibration in the sideband Rabi frequencies or gate time to be $\lesssim 0.1\%$.


\begin{table}
\center
\begin{tabular}{|c|c|c|}\hline
operation 						& error $/10^{-3}$ & ref.\ \\ \hline
memory $(t_g=3.25\ms)/(T_2^*=50\s)$	& 0.07		& \cite{Harty2014}\\
state preparation 					& 0.2			& \cite{Harty2014}\\
global single-qubit gate (benchmarked)		& 0.001		& \cite{Harty2014} \\ 
single-shot readout (per qubit)			& 3			& this work \\ 
two-qubit gate (tomography)			& 3			& this work\\ \hline
\end{tabular}
\caption{%
Summary of errors in elementary qubit operations achieved in the present experimental apparatus for the \Ca{43} (\ket{\up},\ket{\down}) hyperfine ``atomic clock'' qubit. The readout error could be reduced to the $0.5\times10^{-3}$ level measured in~\cite{Harty2014} using spatially-resolved fluorescence detection~\cite{Burrell2010}. Addressed single-qubit gates with $\ish 1\times10^{-3}$ error were demonstrated for \Mg{25} hyperfine qubits using microwave techniques in a similar surface trap by Warring \etal~\cite{Warring2013}.
}
\label{T:summary}
\end{table}

In conclusion, we have introduced a dynamically-decoupled two-qubit gate scheme for trapped-ions, which we have implemented with $99.7\,(1)\%$ fidelity using near-field microwave techniques in a room-temperature microfabricated surface trap. The gate was applied to \Ca{43} hyperfine qubits, for which state-of-the-art single-qubit performance was previously demonstrated in the same apparatus (Table~\ref{T:summary}). Present limits to the gate speed and fidelity are purely technical, and could be improved significantly in future experiments. Heating rates can be decreased using surface cleaning techniques~\cite{Allcock2011,Hite2012} or cryogenic operation~\cite{Labaziewicz2008}. The gate speed could be substantially increased, thereby also reducing its sensitivity to heating and motional mode frequency fluctuations, by moving the ion closer to the trap electrodes or increasing the microwave power. Off-resonant excitation and a.c.\ Zeeman shifts could be reduced by nulling the microwave field using multiple electrodes~\cite{Ospelkaus2008,Warring2013a,AudeCraik2016} or improved trap geometries~\cite{Carsjens2014}. The dynamical decoupling demonstrated here should prove to be  particularly effective when used in combination with these more advanced trap designs, as it significantly reduces the level of field suppression that must be achieved. It may also be useful in mitigating the effects of crosstalk between different trap zones in a multi-zone architecture, arising from the large microwave fields required for multi-qubit gates. Finally, we note the DDMS gate may be useful for laser-driven gates on optical or hyperfine qubits, where a.c.\  Stark shifts can present a significant experimental complication~\cite{Haffner2003}.

Laser-driven two-qubit gates with comparable fidelity have recently been implemented in a surface trap for \ion{Yb}{} hyperfine qubits at Sandia National Laboratories~\cite{Maunz2016}. 

We thank A.~M.~Steane and D.~N.~Stacey, and members of the NIST Ion Storage group, in particular T.~R.~Tan and D.~Slichter, for helpful discussions, and A.~Bermudez for comments on the manuscript. This work was supported by the U.K.\ EPSRC ``Networked Quantum Information Technology'' Hub and the U.S.\ Army Research Office (ref.\ W911NF-14-1-0217).

\section{Supplementary Material} 

\subsection{Microwave fields}

\begin{figure}[t]
\includegraphics[width=0.75\columnwidth]{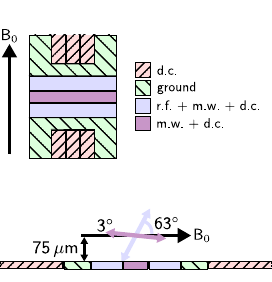}
\caption{Schematic of the trap viewed from above and axially. The ions are confined 75$\,\mu$m above the trap's surface. The $B_0=14.6$\mT{} static field lies in the trap plane, perpendicular to the trap's axis. The two outer axial electrodes carry the r.f.\ trapping voltage. The six radial  electrodes carry d.c.\ voltages which, in combination with the d.c.\ bias applied to the axial electrodes, create the static trapping potential. The three axial electrodes may be driven with microwaves.  The microwave field produced by the right (centre) axial electrode lies at an angle of $63^{\circ}$ ($3.1^{\circ}$) to the B-field~[17,43 \S6-\S6.2].}
\label{fig:trap_schematic}
\end{figure}

Figure \ref{fig:trap_schematic} is a schematic of the trap used in these experiments. It was originally designed to have microwaves applied to all three axial electrodes, allowing the creation of a microwave magnetic field gradient, while maintaining a field null at the trap's centre. However, in this work we take advantage of the DDMS scheme, which allows us to use a simpler ``un-nulled'' configuration,  with the RSB and BSB applied to a single electrode and, correspondingly, a non-zero field amplitude at the trap's centre.

To minimise off-resonant excitation of the $\sigma_+$-polarized qubit transition, we apply the RSB and BSB microwaves to the trap's central axial electrode, which produces a predominantly $\pi$-polarized field (the $\pi$-polarized spectator transitions are detuned by $\approx$50\MHz{} from the qubit), whose gradient has a strong $\sigma$-polarized component. The DDMS carrier drive is applied to the right axial electrode, whose field has a much larger $\sigma$ component (figure 4).

The microwave magnetic field gradient created by each sideband is $\approx$7\,T/m. This gradient is accompanied by a field amplitude of $\approx$780\,$\mu$T (inferred from the measured value of $\Delta$, see below), which drives the qubit ($\pi$ spectator transitions) with a Rabi frequency of 370\,kHz (14\,MHz)~[17,43 \S6-6.2]. We reiterate that, despite these large Rabi frequencies, the data in figure 3 demonstrate that we are able to suppress errors due to off-resonant Rabi flopping (spin flips) to $\lesssim 0.1\%$. 

\begin{table}[b]
\begin{tabularx}{0.75\columnwidth}{cYY}
 & \multicolumn{2}{c}{a.c.\ Zeeman shift (kHz)} \\
Transition & RSB & BSB\\
\hline
$\ket{4,+0} \leftrightarrow \ket{3,-1} $ & $+0.12$ & $+0.13$ \\
$\ket{4,+0} \leftrightarrow \ket{3,+0}$ & $+540$ & $+610$ \\
$\ket{4,+0} \leftrightarrow \ket{3,+1}$ & $+13$   & $-13$ \\
$\ket{4,+1} \leftrightarrow \ket{3,+1}$ & $-600$  & $-530$ \\
$\ket{4,+2} \leftrightarrow \ket{3,+1}$ & $-0.33$ & $-0.31$ \\
\hline
Total & -47 & +68 \\
\end{tabularx}
\caption{a.c.\ Zeeman shifts on the $\ket{4,+0} \leftrightarrow \ket{3,+1}$ qubit transition, arising from off-resonant excitation of the the various ground-level spectator transitions connecting to the qubit states by the RSB and BSB fields. A positive (negative) shift increases (decreases) the qubit frequency. }
\end{table}

\begin{figure*}[t]
  \captionsetup[subfigure]{labelformat=empty}
  \subfloat[]{\includegraphics[width=\columnwidth]{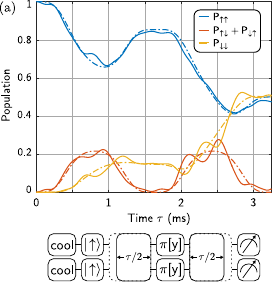}} \hfill
  \subfloat[]{\includegraphics[width=\columnwidth]{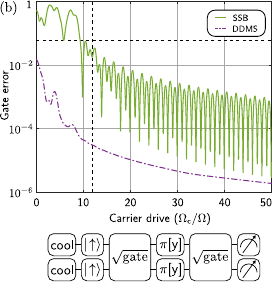}}
  \caption{ Comparison between the dynamically decoupled \MS (DDMS, dash-dot lines) and single sideband (SSB, solid lines) gates.  Unless stated otherwise, we have assumed the parameters used in our experiments: $\Omega = 308\,\mathrm{Hz}\times2\pi$, $\delta = 1.23\,\mathrm{kHz}\times 2\pi$ (a 4-loop gate), $\Omega_{\mathrm{c}} = 3.69\,\mathrm{kHz}\times 2\pi$ and, a $y$-axis $\pi$-pulse mid-way through the gate. (a) Time scan of the spin-state dynamics for both gate schemes. (b) Gate error as a function of carrier Rabi frequency ($\Omega_{\mathrm{c}}$), assuming a constant uncompensated a.c.\ Zeeman shift of $\Delta'=20\,\mathrm{Hz}\times2\pi$. The horizontal dashed line indicates the gate error for a standard MS gate, equivalent to the DDMS gate without the refocussing $\pi$-pulse and with $\Omega_{\mathrm{c}}=0$. The vertical dashed line indicates the value of $\Omega_{\mathrm{c}}$ used in our experiments. }
  \label{fig:time_scans}
\end{figure*}

The a.c.\ Zeeman shifts arising from off-resonant excitation of the various transitions connecting to the qubit states are given in table 2. These values include the Bloch-Siegert correction, which is significant due to the high degree of cancellation between the various shifts. Accordingly, we calculate the shift $\Delta_i$ on the qubit due to off-resonant excitation of a transition with frequency $\omega_i$ by a microwave field with frequency $\omega_{\mathrm{\mu \mathrm{w}}}$ and Rabi frequency on that transition $\Omega_i$ as
\begin{equation}
\Delta_i = \frac{\Omega_i^2 \omega_{i}}{2\left(\omega_i^2-\omega_\mathrm{\mu \mathrm{w}}^2\right)}
 \end{equation}
where $\omega_{\mu \mathrm{w}} = \left(\omega_{0}+\Delta\right) \pm  \left(\omega_{r}+\delta\right)$ for the RSB ($-$) and BSB ($+$), with the values of $\Delta$ and $\omega_{\mathrm{r}}$ used in our experiment (see the main text).

For these calculations, we have assumed the $3.1^\circ$ microwave polarization angle previously measured in~[43]. We determine the overall field amplitude from the measured differential a.c.\ Zeeman shift $\Delta$, assuming equal amplitudes for the RSB and BSB fields. We expect this to give a more accurate estimate of the field than~[43] (based on the technique described in [41]), which would predict a field of 1.0\,mT for our gradient. We do not attempt to include the second-order correction due to the change in $\omega_i$ arising from a.c.\ Zeeman shifts. Instead, we use the bare (un-shifted) values of each $\omega_i$ in our calculations.

\subsection{Comparison between gate schemes}

The simulations shown in figure \ref{fig:time_scans} demonstrate the advantages of the dynamically decoupled \MS (DDMS) gate over the single sideband (SSB) gate. They were performed by numerical integration of the von Neumann equation for the Hamiltonian
\begin{equation}
H = H_{\mathrm{MS}} + H_{\mathrm{c}} + H_{\mathrm{Z}}
\end{equation}
including 2 spin states and a single motional mode, truncated to 30 harmonic oscillator Fock states. For the SSB gate simulations, the BSB Rabi frequency is set to zero, while the RSB Rabi frequency is double that used in the DDMS gate simulations.

Figure \ref{fig:time_scans} (a) shows the spin-state dynamics for both gate schemes. While the DDMS dynamics are identical to those for a standard \MS (MS) gate, the SSB gate is more complicated, featuring characteristic wiggles. These more complicated dynamics are likely to make the SSB gate more difficult to set up and optimize in practice.

Figure \ref{fig:time_scans} (b) shows the gate error as a function of carrier Rabi frequency, assuming a constant uncompensated a.c.\ Zeeman shift of $\Delta'=20\,\mathrm{Hz}\times2\pi$. We see that the DDMS gate does not have a significant error as long as $\Omega_{\mathrm{c}}\gtrsim10\, \Omega$. In contrast, the SSB gate is only able achieve low errors at particular values of $\Omega_{\mathrm{c}}$, making it more difficult to set up and more susceptible to noise in $\Omega_{\mathrm{c}}$. This sensitivity to noise in $\Omega_{\mathrm{c}}$ is likely to mean that a second-order carrier drive is required to achieve high fidelities with the SSB gate. 

We note that the error on the DDMS curve is still lower than that for the standard MS gate when $\Omega_{\mathrm{c}}=0$ ($1.5\,\%$, compared with $6.1\,\%$), demonstrating that the refocussing $\pi$-pulse provides some robustness to a.c.\ Zeeman shifts, even without the carrier drive. This may be useful in experiments with only small uncompensated a.c.\ Zeeman shifts.


\end{document}